# A Quadratic Convex Approximation of Optimal Power Flow in Distribution Systems with Application in Loss Allocation

Tianshu Yang, *Student Member*, *IEEE*, Ye Guo, *Senior Member*, *IEEE*, Lirong Deng, *Student Member*, *IEEE*, Hongbin Sun, *Fellow*, *IEEE*, and Wenchuan Wu, *Senior Member*, *IEEE*

*Abstract*— **In this paper, a novel quadratic convex optimal power flow model, namely, MDOPF, is proposed to determine the optimal dispatches of distributed generators. Based on the results of MDOPF, two price mechanisms, distribution locational marginal price (DLMP) and distribution locational price (DLP), are analyzed. For DLMP, an explicit method is developed to calculate the marginal loss that does not require a backward/forward sweep algorithm and thus reduces the computational complexity. However, the marginal loss component in DLMP will cause over-collection of losses (OCL). To address this issue, DLP is defined, which contains two components, the energy cost component and loss component, where the loss component is determined by the proposed loss allocation method (LAM). Numerical tests show that the proposed MDOPF has a better accuracy than existing OPF models based on linear power flow equations. In addition, the proposed marginal loss method and DLMP algorithm have satisfactory accuracy compared with benchmarks provided by ACOPF, and the proposed DLP can eliminate OCL.**

## I. INTRODUCTION

The increasing penetration of distributed generation (DG) has led to the evolution of active distribution networks (ADNs). Although DG can reduce the cost of electricity supply and establish a diverse energy ecosystem among end-users, it may also cause security hazards without proper dispatch and control. To that end, some studies try to extend optimal power flow (OPF) and pricing mechanisms to the distribution system level.

On the OPF model, due to the nonconvexity of the alternating current OPF (ACOPF) model, it can be computationally intractable even for small systems. Thus, efforts in ACOPF convexification are indispensable for the operation of ADNs. References [1-4] set forth semidefinite programming (SDP) relaxation and prove it will be exact with the allowance of load over-satisfaction and the utilization of virtual phase shifters. However, solutions produced by SDP become physically meaningless when the duality gap is nonzero or the rank-1 solution fails to be obtained [5, 6]. Second-order cone programming (SOCP) relaxation based on the branch flow model [5-7] is proposed and proven to be exact in conditions where the objective function is convex, strictly increasing in branch losses, nonincreasing in loads, and independent of complex branch flows. In addition, SOCP relaxation requires that the voltage upper bounds do not bind at optimality, which is a strict condition and does not apply to systems with a high penetration of DG. Therefore, there is hitherto no relaxation approach that can always provide feasible solutions.

To remedy the shortcomings of convex relaxation, many efforts have focused on linear power flow models, which will naturally make OPF models convex. Existing works include warm-start models (WSMs) [8-11] and cold-start models (CSMs) [11-15]. The former kind of method linearizes the ACPF model around the operating points, thus requiring predetermined initial points, while CSMs do not assume initial operating points. Although there are many studies on linear power flow models, few of them address OPF problems in ADNs. The most classic CSM-based OPF model is the direct current OPF (DCOPF) model. The DCOPF model has good accuracy for transmission network analysis but is not directly applicable to distribution networks [16]. In [17], a non-iterative algorithm is developed to solve the OPF problem in ADNs. They first estimate the optimal operating point of the system with a CSM and then construct an OPF model based on a WSM to improve the accuracy of the solutions. The key to this algorithm is the accuracy of the CSM. If the operating point estimated by the CSM has a large error, the final results of the algorithm may turn out to be suboptimal. Nevertheless, there is no ideal OPF formulation specifically designed for distribution systems.

On the pricing mechanism, the most discussed pricing method is distribution locational marginal price (DLMP) [12, 18], which draws on the experience of locational marginal price (LMP) in wholesale electricity markets. The LMP of a certain location is defined as the marginal cost to supply an increment of load at this location. Therefore, LMP contains three components [19]: an energy cost component, marginal loss component, and congestion component. Due to the rare occurrence of congestion in distribution systems [20], DLMP consists of an energy cost component and marginal loss component. In [21], two approaches, duality and marginal loss, are used to analyze DLMP. Normally, as an OPF model is solved, the DLMP can be obtained by Lagrange multipliers. However, the Lagrange multipliers are inaccurate in the approximate OPF model compared with those in the ACOPF model. To improve the accuracy, a backward/forward sweep algorithm was recently developed in [17] to calculate the marginal loss and LMP.

A related question is whether the marginal loss component in LMP will cause over-collection of losses (OCL) or a loss surplus [27]. Naturally, independent system operators (ISOs) are required to develop methods to redistribute the OCL [28]. In a report [22], it is concluded that ISOs, such as PJM, NYISO, CAISO, ISO-NE and MISO, have no ideal mechanism for dealing with OCL. The essence of collecting the marginal loss and returning the OCL is actually loss allocation. In [23], four methods, a Pro Data allocation method, quadratic allocation method, proportional allocation method and exact method, are proposed to allocate the network losses in radial distribution systems. A branch current decomposition-based approach is proposed in [23, 24] to allocate the network losses for three-phase distribution networks. Although there are many seminal papers presenting loss allocation methods for distribution networks, few studies have examined this issue from the perspective of pricing.

With the above motivation, this paper develops an OPF model, a new DLMP formulation, and a novel loss allocation approach. Therefore, the main contributions of the paper are threefold:

(i) A quadratic convex approximate OPF model, denoted as MDOPF, is proposed for distribution systems. Based on modified DistFlow [25], MDOPF does not require any assumptions related to the R/X ratio or voltage magnitudes and thus has satisfactory accuracy in a wide range of system states. Moreover, in MDOPF, the reactive power is considered, and the thermal limit is modeled as convex inequality constraints, and this approach guarantees the secure operation of ADNs.

(ii) Explicit expressions of the DLMP formulation and marginal loss method (MLM) are proposed for distribution networks and do not require a backward/forward sweep algorithm. In addition, the influence of the bus injected power on the bus voltage magnitude is considered in the MLM, thus improving the accuracy of the DLMP results.

(iii) From the perspective of electricity pricing, we define distribution locational price (DLP) and propose a novel loss allocation method (LAM). DLP consists of two components, an energy cost component and loss component, in which the loss component is determined by the proposed LAM. Moreover, the loss component is decomposed into active power contributions and reactive power contributions. Similar to the proposed MLM, the LAM is expressed in explicit forms.

The remainder of this paper is organized as follows. Section II introduces the modified DistFlow. Section III presents the MDOPF model. Section IV describes the price mechanisms (i.e., the MLM and LAM). Section V outlines the test results of the proposed MDOPF model, the MLM and the LAM using modified IEEE test systems and several larger distribution systems. Section VI concludes the paper.

## II. Preliminary

Modified DistFlow [25] is the latest research on a cold-start linear branch flow (LBF) model for distribution networks and which has satisfactory accuracy compared with other existing cold-start LBF models. Therefore, modified DistFlow is expected to be applied to optimization problems in ADNs.

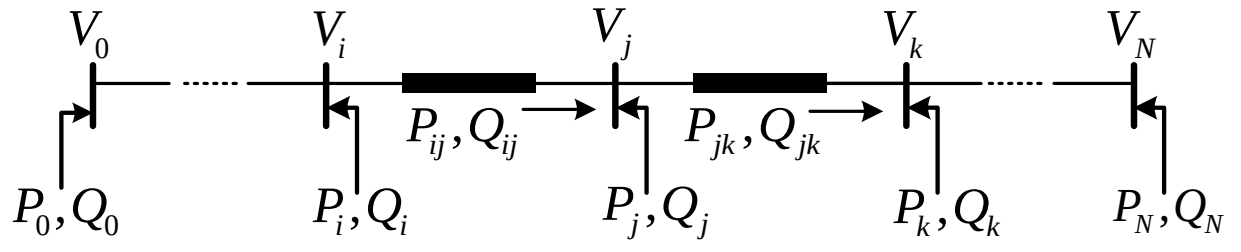

Fig. 1. One-line diagram of a main distribution feeder.

The main idea of modified DistFlow is that the active and reactive power are replaced by their ratios to the voltage magnitude as state variables, so that errors introduced by conventional branch flow linearization approaches due to completely ignoring the quadratic term are reduced.

Consider a single radial network structure shown in Fig. 1. Let $i$ = 0, 1, 2 … N enumerate the buses; then, modified DistFlow can be represented as follows:

$$\hat{P}_{jk} = \hat{P}_{ij} + \hat{P}_j. \tag{1}$$

$$\hat{Q}_{jk} = \hat{Q}_{ij} + \hat{Q}_j. \tag{2}$$

$$W_j - W_i = R_{ij}\hat{P}_{ij} + X_{ij}\hat{Q}_{ij}. \tag{3}$$

$$\hat{P}_i = P_i W_i. \tag{4}$$

$$\hat{Q}_i = Q_i W_i. \tag{5}$$

where $R_{ij}$ and $X_{ij}$ represent the resistance and reactance of branch $ij$, respectively; $P_i$ and $Q_i$ denote the active power injection and reactive power injection at bus $i$, respectively; and $\hat{P}_i$ and $\hat{Q}_i$ are the ratios of $P_i$ and $Q_i$ to the voltage magnitude $V_i$ at bus $i$, respectively. The auxiliary variable $W_i$ is defined as $1/V_i$ and is linearly approximated as:

$$W_i = 2 - V_i. \tag{6}$$

$\hat{P}_{ij}$ and $\hat{Q}_{ij}$ are the ratios of the branch power flows $P_{ij}$ and $Q_{ij}$ to the voltage magnitude $V_i$ on branch $ij$ as follows:

$$\hat{P}_{ij} = P_{ij} W_i. \tag{7}$$

$$\hat{Q}_{ij} = Q_{ij} W_i. \tag{8}$$

For simplicity, we denote $\hat{P}_i$ and $\hat{Q}_i$ as the modified active and reactive power injections, respectively, and $\hat{P}_{ij}$ and $\hat{Q}_{ij}$ as the modified active and reactive branch flows, respectively.

To solve modified DistFlow in an explicit matrix form, a path-branch incidence matrix is introduced in [25]. An entry ($l_{ij}$, $k$) of the path-branch incidence matrix $T$ is defined as:

$$T_{l_{ij},k} = \begin{cases} 1 & l_{ij} \in \Psi_k \\ 0 & l_{ij} \notin \Psi_k \end{cases}. \tag{9}$$

where $l_{ij}$ denotes the branch between buses $i$ and $j$ and $\Psi_k$ is the set of branches on the path of bus $i$.

Then, by incorporating the path-branch incidence matrix, (1) -(3) become:

$$\hat{P}_{Br} = -T\hat{P}_R, \tag{10}$$

$$\hat{Q}_{Br} = -T\hat{Q}_R. \tag{11}$$

$$W_R - W_0 = T^T R_N \hat{P}_{Br} + T^T X_N \hat{Q}_{Br}, \tag{12}$$

where $W_R$ denotes the column vector of $W_i$ for $i$ = 1,2,… N and $\hat{P}_R$ and $\hat{Q}_R$ are the column vectors of $\hat{P}_i$ and $\hat{Q}_i$, respectively. Thereinto, the right subscript $R$ represents the receiving end bus of each branch. $\hat{P}_{Br}$ and $\hat{Q}_{Br}$ represent the column vectors of $\hat{P}_{ij}$ and $\hat{Q}_{ij}$, respectively, and $R_N$ and $X_N$ are the diagonal matrices of $R_{ij}$ and $X_{ij}$, respectively.

Substituting (10) and (11) into (12), we have:

$$W_R = W_0 - T^T R_N T\hat{P}_R - T^T X_N T\hat{Q}_R. \tag{13}$$

Furthermore, substituting (6) into (13), we can obtain:

$$V_R = V_0 + T^T R_N T\hat{P}_R + T^T X_N T\hat{Q}_R, \tag{14}$$

where $V_R$ represents the column vector of $V_i$.

From (13) and (14), we can find that $W_R$ and $V_R$ are affine mappings of $\hat{P}_R$ and $\hat{Q}_R$.

For the PQ load, if the power injections are fixed, according to (4) and (5), (13) and (14) can be solved by:

$$W_R = (I + T^T R_N T P_N + T^T X_N T Q_N)^{-1} W_0. \tag{15}$$

$$V_R = 2 - (I + T^T R_N T P_N + T^T X_N T Q_N)^{-1}(2 - V_0), \tag{16}$$

where $I$ is the identity matrix and $P_N$ and $Q_N$ are the diagonal matrices of $P_i$ and $Q_i$, respectively.

## III. Optimal Power Flow

In this section, we build and convexify an OPF model based on modified DistFlow, denoted as MDOPF.

### A. Model

We assume that the OPF objective function is linear, and minimizes the generation cost. Therefore, the original OPF model is:

$$\text{OPF: } \min_{\substack{V_i, P_i^G, Q_i^G, \\ P_{ij}, Q_{ij}.}} : \sum_{i\in\Phi_G} C_i^P P_i^G + C_i^Q Q_i^G, \tag{17}$$

$s.t.$

$$\sum_{j:i\to j} P_{ij} = P_i + \sum_{h:h\to i}\left(P_{hi} - R_{hi}\frac{P_{hi}^2 + Q_{hi}^2}{V_h^2}\right), \forall i \in \Phi_B, \tag{18}$$

$$\sum_{j:i\to j} Q_{ij} = Q_i + \sum_{h:h\to i}\left(Q_{hi} - X_{hi}\frac{P_{hi}^2 + Q_{hi}^2}{V_h^2}\right), \forall i \in \Phi_B, \tag{19}$$

$$V_i^2 - V_j^2 = R_{ij}P_{ij} + X_{ij}Q_{ij} - \left(R_{ij}^2 + X_{ij}^2\right)\frac{P_{hi}^2 + Q_{hi}^2}{V_h^2}, \forall ij \in \Phi_L, \tag{20}$$

$$P_i = -P_i^D + P_i^G, \forall i \in \Phi_B, \tag{21}$$

$$Q_i = -Q_i^D + Q_i^G, \forall i \in \Phi_B, \tag{22}$$

$$\underline{P}_i^G \le P_i^G \le \overline{P}_i^G, \forall i \in \Phi_G, \tag{23}$$

$$\underline{Q}_i^G \le Q_i^G \le \overline{Q}_i^G, \forall i \in \Phi_G, \tag{24}$$

$$\underline{V}_i \le V_i \le \overline{V}_i, \forall i \in \Phi_B, \tag{25}$$

$$P_{ij}^2 + Q_{ij}^2 \le \overline{S}_{ij}^2, \forall ij \in \Phi_L, \tag{26}$$

where $\Phi_B$ is the set of all buses, $\Phi_L$ is the set of all branches, $\Phi_G$ is the set of all DG buses, $P_i^G$ and $Q_i^G$ are the active and reactive power generation provided by DG, respectively, $P_i^D$ and $Q_i^D$ are the active and reactive power loads, respectively, $\overline{V}_i$ and $\underline{V}_i$ are the upper bound and lower bound of the voltage magnitude, respectively, $\overline{P}_i^G$ and $\underline{P}_i^G$ are the upper bound and lower bound of active power generation of DG, respectively, $\overline{Q}_i^G$ and $\underline{Q}_i^G$ are the upper bound and lower bound of reactive power generation of DG, respectively, $\overline{S}_{ij}^2$ is the capacity of branch $ij$, and $C_i^P$ and $C_i^Q$ are the active and reactive power costs of DG, respectively.

Since the variables in modified DistFlow are $\hat{P}_i$, $\hat{Q}_i$, $\hat{P}_{ij}$ and $\hat{Q}_{ij}$, to construct MDOPF, the objective function (17) should be transformed to:

$$\sum_{i\in\Phi_G} V_i C_i^P \hat{P}_i^G + V_i C_i^Q \hat{Q}_i^G. \tag{27}$$

Then, the branch flow equations (18) and (19) become:

$$\sum_{j\in \mathrm{N}_i} \hat{P}_{ij} + \hat{P}_i = 0, \forall i \in \Phi_B, \tag{28}$$

$$\sum_{j\in \mathrm{N}_i} \hat{Q}_{ij} + \hat{Q}_i = 0, \forall i \in \Phi_B, \tag{29}$$

where $\mathrm{N}_i$ represents the sets of neighboring buses of bus $i$.

The voltage equation (20) becomes:

$$W_j - W_i = R_{ij}\hat{P}_{ij} + X_{ij}\hat{Q}_{ij}, \forall ij \in \Phi_L, \tag{30}$$

$$V_i = 2 - W_i, \forall i \in \Phi_B, \tag{31}$$

The active and reactive power injection constraints (21) and (22) become:

$$\hat{P}_i = -P_i^D W_i + \hat{P}_i^G, \forall i \in \Phi_B, \tag{32}$$

$$\hat{Q}_i = -Q_i^D W_i + \hat{Q}_i^G, \forall i \in \Phi_B, \tag{33}$$

where $\hat{P}_i^G$ and $\hat{Q}_i^G$ are the ratios of the power generation $P_i^G$ and $Q_i^G$ to the voltage magnitude $V_i$ at bus $i$. For simplicity, we denote $\hat{P}_i^G$ and $\hat{Q}_i^G$ as the modified active and reactive power generation, respectively. After MDOPF is solved, $P_i^G$ and $Q_i^G$ can be recovered from:

$$P_i^G = \frac{\hat{P}_i^G}{W_i}. \tag{34}$$

$$Q_i^G = \frac{\hat{Q}_i^G}{W_i}. \tag{35}$$

The DG operation constraints (23) and (24) become:

$$\underline{P}_i^G W_i \le \hat{P}_i^G \le \overline{P}_i^G W_i, \forall i \in \Phi_G, \tag{36}$$

$$\underline{Q}_i^G W_i \le \hat{Q}_i^G \le \overline{Q}_i^G W_i, \forall i \in \Phi_G, \tag{37}$$

The bus voltage limit constraint (25) becomes:

$$2 - \overline{V}_i \le W_i \le 2 - \underline{V}_i, \forall i \in \Phi_B, \tag{38}$$

The branch flow limit (26) becomes the thermal limit as follows:

$$\hat{P}_{ij}^2 + \hat{Q}_{ij}^2 \le \overline{I}_{ij}^2, \forall ij \in \Phi_L, \tag{39}$$

where $\overline{I}_{ij}^2$ represents the upper bound of the current on branch $ij$.

In summary, MDOPF is:

$$\text{MDOPF: } \min_{\substack{W_i, V_i, \hat{P}_{ij}, \hat{Q}_{ij}, \\ \hat{P}_i, \hat{Q}_i, \hat{P}_i^G, \hat{Q}_i^G.}} : \sum_{i\in\Phi_G} V_i C_i^P \hat{P}_i^G + V_i C_i^Q \hat{Q}_i^G, \tag{40}$$

$s.t.$: (28)-(33), (36)-(39)

The decision variables herein are the modified power generation $\hat{P}_i^G$ and $\hat{Q}_i^G$ of DG, modified power injections $\hat{P}_i$ and $\hat{Q}_i$, voltage magnitude $V_i$ and its auxiliary variable $W_i$, and modified branch flows $\hat{P}_{ij}$ and $\hat{Q}_{ij}$.

Note that all the constraints in MDOPF are either linear or convex, while the objective function contains bilinear terms, which are nonconvex.

### B. Convexification

To transform (40) into a convex objective function, the objective should be first written in rectangular representation as follows:

$$\min V^T C^P \hat{P}^G + V^T C^Q \hat{Q}^G. \quad (41)$$

where $\hat{P}^G = [\hat{P}_0^G; \hat{P}_R^G]$ and $\hat{Q}^G = [\hat{Q}_0^G; \hat{Q}_R^G]$. Here, $\hat{P}_R^G$ and $\hat{Q}_R^G$ are vectors of $\hat{P}_i^G$ and $\hat{Q}_i^G$, respectively, for $i$ = 1,2, …, N. $C^P$ and $C^Q$ are the active and reactive power prices of generators, respectively, and can be decomposed as follows:

$$C^P = \begin{bmatrix} C_0^P & 0 \\ 0 & C_N^p \end{bmatrix}, C^Q = \begin{bmatrix} C_0^Q & 0 \\ 0 & C_N^Q \end{bmatrix}, \quad (42)$$

where $C_N^p$ and $C_N^Q$ are the diagonal matrices of $C_i^P$ and $C_i^Q$, respectively, for $i$ = 1,2, …, N.

Then, $V$ should be replaced by $\hat{P}$ and $\hat{Q}$. Let $V$ denote the voltage magnitudes of all buses, and $V$ can be decomposed into $V = [V_0; V_R]$, where $V_0$ is the voltage magnitude of the slack bus.

$\hat{P}_R$ and $\hat{Q}_R$ are defined as:

$$\hat{P}_R = -P_N^D W_R + \hat{P}_R^G. \quad (43)$$

$$\hat{P}_R = -P_N^D W_R + \hat{P}_R^G, \quad (44)$$

where $P_N^D$ and $Q_N^D$ are diagonal matrices of $P_i^D$ and $Q_i^D$, respectively, for $i$ = 1,2, …, N.

According to (14), $V_R$ can be written as follows:

$$V_R = V_R^D + T^T R_N T \hat{P}_R^G + T^T X_N T \hat{Q}_R^G, \quad (45)$$

where $V_R^D$ is obtained by:

$$V_R^D = V_0 - T^T R_N T P_N^D W_R - T^T X_N T Q_N^D W_R, \quad (46)$$

According to (6), $V_R^D$ can be solved as follows:

$$V_R^D = \mathbf{2} - (I - T^T R_N T P_N^D - T^T X_N T Q_N^D)^{-1}(2 - V_0). \quad (47)$$

(45)-(47) show that if $P_N^D$ and $Q_N^D$ are given, $V_R^D$ is fixed, and $V_R$ is an affine mapping of $\hat{P}_R^G$ and $\hat{Q}_R^G$.

Next, $V_R$ is replaced by $\hat{P}_R^G$ and $\hat{Q}_R^G$, and the objective function can be transformed into a quadratic function. For clarity, the objectives are split into three parts as follows:

$$\min C_1(\hat{P}_0^G, \hat{Q}_0^G) + C_2(\hat{P}_R^G, \hat{Q}_R^G) + C_3(\hat{P}_R^G, \hat{Q}_R^G), \quad (48)$$

where $C_1(\hat{P}_0^G, \hat{Q}_0^G)$ is the total cost of DG at the slack bus as follows:

$$C_1(\hat{P}_0^G, \hat{Q}_0^G) = V_0 C_0^P \hat{P}_0^G + V_0 C_0^Q \hat{Q}_0^G \quad (49)$$

$C_2(\hat{P}_R^G, \hat{Q}_R^G)$ and $C_3(\hat{P}_R^G, \hat{Q}_R^G)$ represent the total costs of DG at other locations. According to (41) and (45), $C_2(\hat{P}_R^G, \hat{Q}_R^G)$ and $C_3(\hat{P}_R^G, \hat{Q}_R^G)$ are obtained as follows:

$$C_2(\hat{P}_R^G, \hat{Q}_R^G) = \left(V_R^D\right)^T C_N^P \hat{P}_R^G + \left(V_R^D\right)^T C_N^Q \hat{Q}_R^G, \quad (50)$$

$$\begin{aligned} C_3(\hat{P}_R^G, \hat{Q}_R^G) &= \left(T^T R_N T \hat{P}_R^G\right)^T C_N^P \hat{P}_R^G + \left(T^T X_N T \hat{Q}_R^G\right)^T C_N^Q \hat{P}_R^G \\ &+ \left(T^T R_N T \hat{P}_R^G\right)^T C_N^P \hat{Q}_R^G + \left(T^T X_N T \hat{Q}_R^G\right)^T C_N^Q \hat{Q}_R^G. \end{aligned} \quad (51)$$

As the objective function (41) is transformed into a quadratic function (48), we need to prove that the quadratic function is convex.

*Theorem*: If the summation of the trace of $C_N^P$ and the trace of $C_N^Q$ is larger than zero, the quadratic function (48) is convex.

*Proof*: See appendix. ■

In conclusion, the objective function (48) is a convex quadratic function, and the OPF problem is formulated as quadratic programming.

## IV. Price Mechanisms

To meet the demand in a real power system, power transmission will cause network losses. In this section, we first develop the MLM and present an algorithm to calculate DLMP. Then, we define DLP based on the proposed LAM.

### A. Network loss

The total network loss can be obtained by:

$$Pl = \sum_{ij \in \Phi_L} R_{ij} \frac{P_{ij}^2 + Q_{ij}^2}{V_i^2} = \sum_{ij \in \Phi_L} R_{ij} \hat{P}_{ij}^{\ 2} + \sum_{ij \in \Phi_L} R_{ij} \hat{Q}_{ij}^{\ 2}. \quad (52)$$

$$Ql = \sum_{ij \in \Phi_L} X_{ij} \frac{P_{ij}^2 + Q_{ij}^2}{V_i^2} = \sum_{ij \in \Phi_L} X_{ij} \hat{P}_{ij}^{\ 2} + \sum_{ij \in \Phi_L} X_{ij} \hat{Q}_{ij}^{\ 2}. \quad (53)$$

where $Pl$ and $Ql$ denote the total active power loss and total reactive power loss, respectively.

According to (10) and (11), (52) and (53) can be written in rectangular form as follows:

$$Pl = (T\hat{P}_R)^T R_N T \hat{P}_R + (T\hat{Q}_R)^T R_N T \hat{Q}_R. \quad (54)$$

$$Ql = (T\hat{P}_R)^T X_N T \hat{P}_R + (T\hat{Q}_R)^T X_N T \hat{Q}_R. \quad (55)$$

In (54) and (55), the total network loss consists of two parts, and each part is a complete square formula of the modified power injection with a coefficient $R$ or $X$. Therefore, the network loss can be decomposed into active power contributions and reactive power contributions.

$$Pl^P = (T\hat{P}_R)^T R_N T \hat{P}_R. \quad (56)$$

$$Pl^Q = (T\hat{Q}_R)^T R_N T \hat{Q}_R. \quad (57)$$

$$Ql^P = (T\hat{P}_R)^T X_N T \hat{P}_R. \quad (58)$$

$$Ql^Q = (T\hat{Q}_R)^T X_N T \hat{Q}_R. \quad (59)$$

### B. DLMP

After MDOPF is solved, DLMP can be obtained by the following definition:

$$DLMP_i^P = \frac{\lambda_i^P}{V_i}, DLMP_i^Q = \frac{\lambda_i^Q}{V_i}. \quad (60)$$

where $\lambda_i^P$ and $\lambda_i^Q$ are the shadow prices in constraints (32) and (33). $DLMP_i^P$ and $DLMP_i^Q$ represent the active power price and reactive power price, respectively, at bus $i$. However, because modified DistFlow is an approximation of the power flow, the accuracy is not high enough if the dual variables are used to calculate DLMP. Therefore, we propose an accurate method to calculate DLMP.

Assume that there is no congestion in the system and that the generation at the power supply point (PSP) is always larger than zero. According to the definition of LMP, which contains only an energy component and marginal loss component, DLMP can be obtained by:

$$DLMP_i^P = C_0^P - C_0^P \cdot \frac{\partial Pl}{\partial P_i} - C_0^Q \cdot \frac{\partial Ql}{\partial P_i}. \quad (61)$$

$$DLMP_i^Q = C_0^Q - C_0^P \cdot \frac{\partial Pl}{\partial Q_i} - C_0^Q \cdot \frac{\partial Ql}{\partial Q_i}. \quad (62)$$

Then, the loss factors w.r.t. the total system losses can be calculated by the following four equations:

$$\frac{\partial Pl}{\partial P_i} = 2\left(T\frac{\partial \hat{P}_R}{\partial P_i}\right)^T R_N T\hat{P}_R + 2\left(T\frac{\partial \hat{Q}_R}{\partial P_i}\right)^T R_N T\hat{Q}_R \quad (63)$$

$$\frac{\partial Pl}{\partial Q_i} = 2\left(T\frac{\partial \hat{P}_R}{\partial Q_i}\right)^T R_N T\hat{P}_R + 2\left(T\frac{\partial \hat{Q}_R}{\partial Q_i}\right)^T R_N T\hat{Q}_R \quad (64)$$

$$\frac{\partial Ql}{\partial P_i} = 2\left(T\frac{\partial \hat{P}_R}{\partial P_i}\right)^T X_N T\hat{P}_R + 2\left(T\frac{\partial \hat{Q}_R}{\partial P_i}\right)^T X_N T\hat{Q}_R \quad (65)$$

$$\frac{\partial Ql}{\partial Q_i} = 2\left(T\frac{\partial \hat{P}_R}{\partial Q_i}\right)^T X_N T\hat{P}_R + 2\left(T\frac{\partial \hat{Q}_R}{\partial Q_i}\right)^T X_N T\hat{Q}_R \quad (66)$$

Note that $W_i$ is defined as $1/V_i$. To obtain more accurate results, the partial derivatives of $\hat{P}_i$ and $\hat{Q}_i$ to $P_i$ and $Q_i$ are:

$$\frac{\partial \hat{P}_i}{\partial P_j} = \begin{cases} -\dfrac{P_i}{V_i^2}\cdot\dfrac{\partial V_i}{\partial P_j} & , i \neq j \\ -\dfrac{1}{V_i} - \dfrac{P_i}{V_i^2}\cdot\dfrac{\partial V_i}{\partial P_j} & , i = j \end{cases} \quad (67)$$

$$\frac{\partial \hat{P}_i}{\partial Q_j} = -\frac{P_i}{V_i^2}\cdot\frac{\partial V_i}{\partial Q_j} \quad (68)$$

$$\frac{\partial \hat{Q}_i}{\partial P_j} = -\frac{Q_i}{V_i^2}\cdot\frac{\partial V_i}{\partial P_j} \quad (69)$$

$$\frac{\partial \hat{Q}_i}{\partial Q_j} = \begin{cases} -\dfrac{Q_i}{V_i^2}\cdot\dfrac{\partial V_i}{\partial Q_j} & , i \neq j \\ -\dfrac{1}{V_i} - \dfrac{Q_i}{V_i^2}\cdot\dfrac{\partial V_i}{\partial Q_j} & , i = j \end{cases} \quad (70)$$

The partial derivative of voltage to injected power can be obtained from the Jacobi matrix. Let $J$ denote the Jacobi matrix:

$$J = \begin{bmatrix} \dfrac{\partial P}{\partial \delta} & \dfrac{\partial P}{\partial V} \\ \dfrac{\partial Q}{\partial \delta} & \dfrac{\partial Q}{\partial V} \end{bmatrix} \quad (71)$$

$$\frac{\partial V}{\partial P} = \left(\frac{\partial Q}{\partial V}\right)^{-1}\frac{\partial Q}{\partial \delta}\left(\frac{\partial P}{\partial \delta} - \frac{\partial P}{\partial V}\left(\frac{\partial Q}{\partial V}\right)^{-1}\frac{\partial Q}{\partial \delta}\right)^{-1} \quad (72)$$

$$\frac{\partial V}{\partial Q} = -\left(\frac{\partial Q}{\partial V} - \frac{\partial Q}{\partial \delta}\left(\frac{\partial P}{\partial \delta}\right)^{-1}\frac{\partial P}{\partial V}\right)^{-1} \quad (73)$$

where $\delta$ denotes the phase angle. As mentioned in [25], $\delta$ can be calculated according to:

$$V_j \sin\delta_{ij} = X_{ij}\hat{P}_{ij} - R_{ij}\hat{Q}_{ij}. \quad (74)$$

For clarity, the steps to calculate DLMP are as follows:

**Algorithm 1**

**procedure** MDOPF
    $V^D \leftarrow V_0$, $\{T, R_N, X_N, P_N^D, Q_N^D\}$
    Construct MDOPF: $F(V_i, W_i, \hat{P}_i^G, \hat{Q}_i^G, \hat{P}_{ij}, \hat{Q}_{ij})$
    $\{V_i^*, W_i^*, \hat{P}_i^{G*}, \hat{Q}_i^{G*}, \hat{P}_{ij}^*, \hat{Q}_{ij}^*\} = \arg\min F(V_i, W_i, \hat{P}_i^G, \hat{Q}_i^G, \hat{P}_{ij}, \hat{Q}_{ij})$
    $\{\delta_i\} \leftarrow \sin\delta_{ij} = (X_{ij}\hat{P}_{ij}^* - R_{ij}\hat{Q}_{ij}^*)/V_j$
    $\{\frac{\partial V}{\partial P}, \frac{\partial V}{\partial Q}\} \leftarrow J(V_i^*, \delta_i)$
    Calculate $\{\frac{\partial \hat{P}_i}{\partial P_j}, \frac{\partial \hat{P}_i}{\partial Q_j}, \frac{\partial \hat{Q}_i}{\partial P_j}, \frac{\partial \hat{Q}_i}{\partial Q_j}\}$ by (67)-(70)
    Calculate $\{\frac{\partial Pl}{\partial P_i}, \frac{\partial Pl}{\partial Q_i}, \frac{\partial Ql}{\partial P_i}, \frac{\partial Ql}{\partial Q_i}\}$ by (63)-(66)
    Calculate $\{DLMP_i^P, DLMP_i^Q\}$ by (61) and (62)
**end procedure**

Although the MLM is widely adopted, there is a problem. Fig. 2 shows the differences between the real network losses and the losses collected by the MLM. It can be seen that the MLM will cause OCL [27].

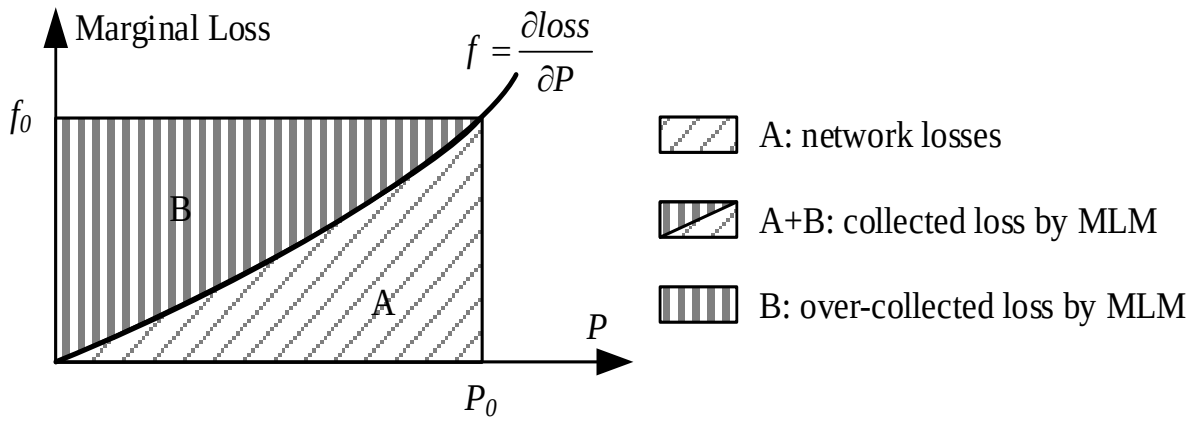

Fig. 2. Marginal loss curve.

In this paper, we assume that there is no congestion. Then, the OCL can be obtained by:

$$OCL = Revenue - Payment. \quad (75)$$

## C. DLP

To return the over-collected revenue, in this subsection, we define DLP, which contains two components, an energy cost component and loss component, as follows:

$$DLP_i^P = C_0^P - C_0^P \cdot \frac{Pl_i^P}{P_i} - C_0^Q \cdot \frac{Ql_i^P}{P_i}. \quad (76)$$

$$DLP_i^Q = C_0^Q - C_0^P \cdot \frac{Pl_i^Q}{Q_i} - C_0^Q \cdot \frac{Ql_i^Q}{Q_i}. \quad (77)$$

In contrast to the loss factor, which considers the incremental changes in network losses due to bus power injections, the total network losses allocated to bus $k$, i.e., $Pl_k^P$, $Ql_k^P$, $Pl_k^Q$, and $Ql_k^Q$, are the summation of the branch loss allocation results as follows:

$$Pl_k^P = \sum_{l_{ij}\in\Psi_k} Pl_{ij,k}^P, \quad (78)$$

$$Ql_k^P = \sum_{l_{ij}\in\Psi_k} Ql_{ij,k}^P, \quad (79)$$

$$Pl_k^Q = \sum_{l_{ij}\in\Psi_k} Pl_{ij,k}^Q, \quad (80)$$

$$Ql_k^Q = \sum_{l_{ij}\in\Psi_k} Ql_{ij,k}^Q. \quad (81)$$

The branch losses are allocated to the related buses. Specifically, if the network loss on branch $ij$ should be allocated to bus $k$, it follows the following rules:

- According to the path-branch incidence matrix, branch $ij$ should be on the path from bus $k$ to the slack bus.
- The amount of the branch loss allocated to bus $k$ is linearly related to the modified branch flows $\hat{P}_{ij}$ and $\hat{Q}_{ij}$ and to the branch impedance $R_{ij}$ and $X_{ij}$.
- The amount of the branch loss allocated to bus $k$ is linearly related to the modified power injections $\hat{P}_k$ and $\hat{Q}_k$.

Combining (10) and (11), the loss allocation results $Pl_{ij,k}^{P}$, $Ql_{ij,k}^{P}$, $Pl_{ij,k}^{Q}$ and $Ql_{ij,k}^{Q}$ are defined as:

$$Pl_{ij,k}^{P} = \begin{cases} \hat{P}_k R_{ij} T_{ij} \hat{P}_R & l_{ij} \in \Psi_k \\ 0 & l_{ij} \notin \Psi_k \end{cases}, \tag{82}$$

$$Ql_{ij,k}^{P} = \begin{cases} \hat{P}_k X_{ij} T_{ij} \hat{P}_R & l_{ij} \in \Psi_k \\ 0 & l_{ij} \notin \Psi_k \end{cases}, \tag{83}$$

$$Pl_{ij,k}^{Q} = \begin{cases} \hat{Q}_k R_{ij} T_{ij} \hat{Q}_R & l_{ij} \in \Psi_k \\ 0 & l_{ij} \notin \Psi_k \end{cases}, \tag{84}$$

$$Ql_{ij,k}^{Q} = \begin{cases} \hat{Q}_k X_{ij} T_{ij} \hat{Q}_R & l_{ij} \in \Psi_k \\ 0 & l_{ij} \notin \Psi_k \end{cases}. \tag{85}$$

Substituting (82)-(85) into (78)-(81):

$$Pl_k^P = T_k^T \hat{P}_k R_N T \hat{P}_R. \tag{86}$$

$$Ql_k^P = T_k^T \hat{P}_k X_N T \hat{P}_R. \tag{87}$$

$$Pl_k^Q = T_k^T \hat{Q}_k R_N T \hat{Q}_R. \tag{88}$$

$$Ql_k^Q = T_k^T \hat{Q}_k X_N T \hat{Q}_R. \tag{89}$$

Then, DLP becomes:

$$DLP_i^P = C_0^P - C_0^P T_i^T V_i^{-1} R_N T \hat{P}_R - C_0^Q T_i^T V_i^{-1} X_N T \hat{P}_R. \tag{90}$$

$$DLP_i^Q = C_0^Q - C_0^P T_i^T V_i^{-1} R_N T \hat{Q}_R - C_0^Q T_i^T V_i^{-1} X_N T \hat{Q}_R. \tag{91}$$

Additionally, DLP can be obtained in a noniterative manner as follows:

$$DLP_R^P = C_0^P - C_0^P (TV_N^{-1})^T R_N T \hat{P}_R - C_0^Q (TV_N^{-1})^T X_N T \hat{P}_R. \tag{92}$$

$$DLP_R^Q = C_0^Q - C_0^P (TV_N^{-1})^T R_N T \hat{Q}_R - C_0^Q (TV_N^{-1})^T X_N T \hat{Q}_R. \tag{93}$$

## V. Numerical Tests

In this section, many scenarios based on a 33-bus system [7], 69-bus system [23] and 141-bus system are set to test the proposed MDOPF model and price mechanisms. In these scenarios, MDOPF is applied to determine the optimal dispatch, and then, the proposed pricing method is implemented. The voltage of the PSP is set to 1.05 p.u., and the generation costs at the PSP are set to 30 \$/MWh and 3 \$/MVarh for active and reactive power, respectively.

The benchmarking ACOPF results are calculated with MATPOWER. MDOPF is solved by an embedded IBM CPLEX 12.8 solver with the YALMIP interface. All the simulations are programmed in MATLAB on a laptop with an Intel Core i7-5600U 2.60 GHz CPU and 8 GB of RAM.

### A. Dispatch results on the 33-bus system

To compare the optimal dispatch of MDOPF and LOPF-D, seven scenarios were set. In each scenario, there was one DG with a capacity of (1 MW, 0.5 MVar). Its reactive power cost was set to 2 \$/MVarh, and its active power cost and location information are shown in Table I. (Note that to keep the DG's active output as a marginal unit, the active power cost of DG was set slightly higher than that of the PSP). Then, we used ACOPF, MDOPF and LOPF-D to calculate the OPF results and showed the generation costs and dispatches. Since all the reactive power dispatches were 0.5 MVar, they are not shown in the table.

The table shows that MDOPF generates more accurate optimal values and solutions than ACOPF. For scenarios 1, 2, 5 and 6, the results of ACOPF and MDOPF are very close. Although there are some differences in the dispatch results of ACOPF and MDOPF in scenarios 3, 4 and 7, their generation costs are only slightly different. The total generation cost determined by LOPF-D is always larger than that of MDOPF.

TABLE I
Optimal Dispatch Comparison

| No. | DG Location | Price (\$/MWh) | Generation Cost (\$) | | | PG of DG (MW) | | |
|---|---|---|---|---|---|---|---|---|
| | | | ACOPF | MDOPF | LOPF-D | ACOPF | MDOPF | LOPF-D |
| 1 | 18th bus | 31 | 122.16 | 122.16 | 122.45 | 0.625 | 0.624 | 1.000 |
| 2 | 25th bus | 31 | 123.32 | 123.32 | 123.55 | 0.353 | 0.368 | 1.000 |
| 3 | 33rd bus | 31 | 121.62 | 121.66 | 121.66 | 0.822 | 1.000 | 1.000 |
| 4 | 6th bus | 32 | 122.96 | 123.00 | 123.23 | 0.233 | 0.513 | 1.000 |
| 5 | 12th bus | 32 | 122.57 | 122.58 | 122.85 | 0.515 | 0.614 | 1.000 |
| 6 | 15th bus | 32 | 122.53 | 122.53 | 122.99 | 0.478 | 0.502 | 1.000 |
| 7 | 31st bus | 32 | 122.23 | 122.28 | 122.54 | 0.501 | 0.704 | 1.000 |

### B. DLMP results on the 33-bus system

In this subsection, the proposed DLMP algorithm was compared with the LOPF-D algorithm mentioned in [17]. Therefore, the following four scenarios were set the same as those in [17].

*Scenario A1*: There were 4 identical DGs installed at Buses 18, 22, 25, and 33, each with an output range of [0, 0.2] MW and [0, 0.1] MVar. The real power price of DG was set to \$31/MWh, which was \$1/MWh higher than that at the PSP, and the reactive power price was set to \$4/MVarh, which was \$1/MVarh higher than that at the PSP.

*Scenario A2*: (**High DG penetration scenario**) The prices of DG were set to \$25/MWh and \$2/MVarh, which were both lower than those at the PSP. The size of each DG was increased to [0, 1] MW and [0, 0.5] MVar. In addition, a load of 0.5 MW was added to the PSP, representing the load from the transmission level to create reverse flow.

*Scenario A3*: (**Heavy load scenario**) The load of the 33-bus system was scaled up to 150% of the baseload, and no DG connection was considered.

*Scenario A4*: (**High-impedance scenario**) The impedance of each branch in the 33-bus system was increased by 190%, and no DG connection was considered.

The detailed results of DLMP and the errors w.r.t. ACOPF are shown in Fig. 3 and Fig. 4. A summary of the average errors of the proposed method compared with those of the method in [17] is shown in Table II. Note that the DLMP results of the proposed method are marked as MDOPF. Because the optimal dispatch results of the two methods are identical, which can also be reflected in the accuracy of DLMP, the dispatch results are not demonstrated.

Fig. 3 shows that the active power prices calculated by the proposed method are very close to the benchmarks determined by ACOPF. In scenarios A1 and A2, the systems work at the baseload of the 33-bus system with different penetrations of DG. The errors in A1 and A2 are almost the same, which shows that the DG output has little effect on the accuracy. In scenarios A3 and A4, as the load/impedance increases, the DLMP errors of the buses located at the end of the feeders become larger. However, the voltage magnitudes of the buses decrease to 0.87 p.u.~0.9 p.u., which means that extreme conditions happen. Such low voltage magnitudes would not be allowed in real operating conditions but are used only to demonstrate the performance of the proposed approach here. Even in such conditions, the DLMP errors of MDOPF are negligible.

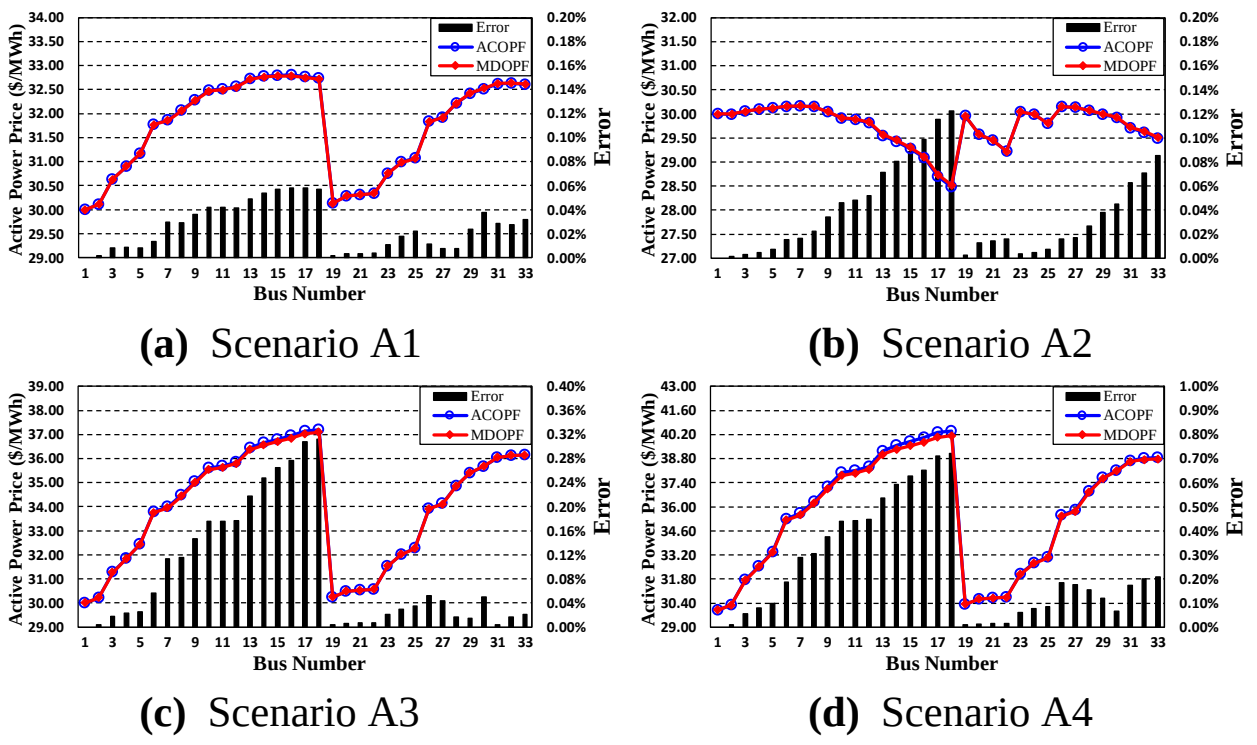

(a) Scenario A1 (b) Scenario A2

(c) Scenario A3 (d) Scenario A4

Fig. 3. Active power prices.

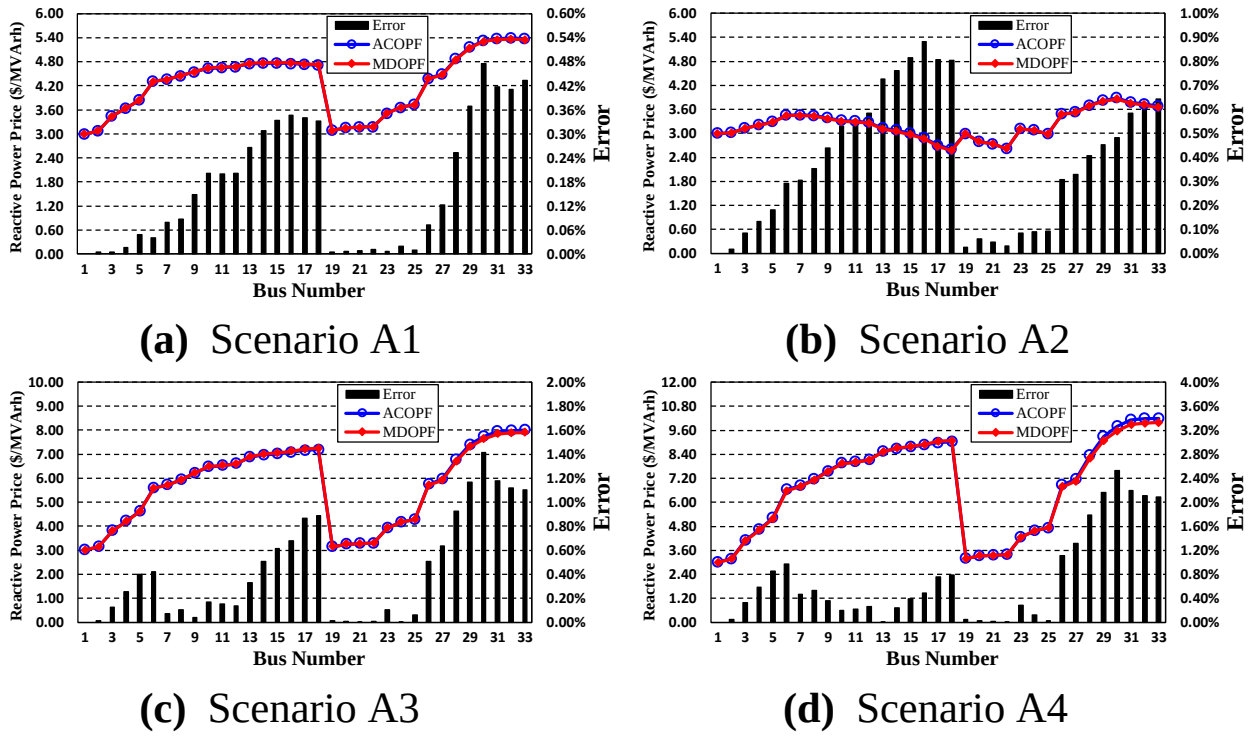

(a) Scenario A1 (b) Scenario A2

(c) Scenario A3 (d) Scenario A4

Fig. 4. Reactive power prices.

As shown in Fig. 4, the reactive power prices follow a similar pattern to the active power prices. However, the errors are larger than those of the active power prices. This is because the reactive power cost is 3 $/MVarh, which is much lower than the active power cost.

TABLE II
SUMMARY OF AVERAGE ERRORS OF DLMP RESULT

| Scenarios | | A1 | A2 | A3 | A4 |
|---|---|---|---|---|---|
| Active Price | MDOPF | 0.02% | 0.04% | 0.09% | 0.25% |
| | LOPF-D | 0.18% | 0.08% | 0.97% | 1.96% |
| Reactive Price | MDOPF | 0.17% | 0.39% | 0.44% | 0.73% |
| | LOPF-D | 0.31% | 0.61% | 3.34% | 5.84% |

The table shows that the proposed method yields more accurate DLMPs than LOPF-D, especially in scenarios A3 and A4, where MDOPF shows more merits. This is because LOPF-D is a WSM and thus requires an accurate operating point. As the load/impedance of the system increase, LPF-D [17] can hardly provide accurate operating points, resulting in a significant rise in DLMP errors. Instead, MDOPF can obtain more precise power flow solutions, and thus, the obtained DLMPs are more reliable.

## C. Large active distribution networks

To verify the proposed MDOPF, MLM and LAM in different systems, three large ADNs were tested. The 33-bus system, 69-bus system, and 141-bus system were used as the basic systems, and we used the method mentioned in [26] to extend those systems to large systems. The details are described as follows:

*Scenario C1*: A *3201-bus system* was obtained by duplicating the 33-bus system 100 times. There were 400 DGs distributed at the end of the distribution feeders.

*Scenario C2*: A *6801-bus system* was obtained by duplicating the 69-bus system 100 times. There were 400 DGs distributed at the end of the distribution feeders.

*Scenario C3*: A *14001-bus system* was obtained by duplicating the 141-bus system 100 times. There were 600 DGs distributed at the end of the distribution feeders.

The capacity of all DGs was set to [0.2 MW, 0.1 MVar], and the biddings of those DGs were 25 $/MWh and 2 $/MVarh for active and reactive power, respectively. In addition, the above three systems were modified by randomly scaling individual branch impedances ($R_{ij}$, $X_{ij}$) and loads ($P_i^D$, $Q_i^D$) in the range of (0.7, 1.3), respectively.

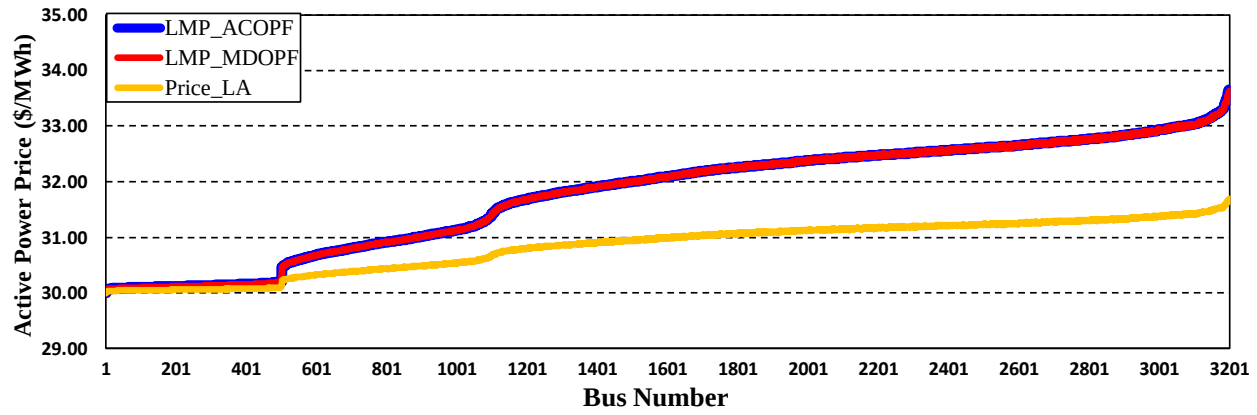

Fig. 6. 3201-bus system.

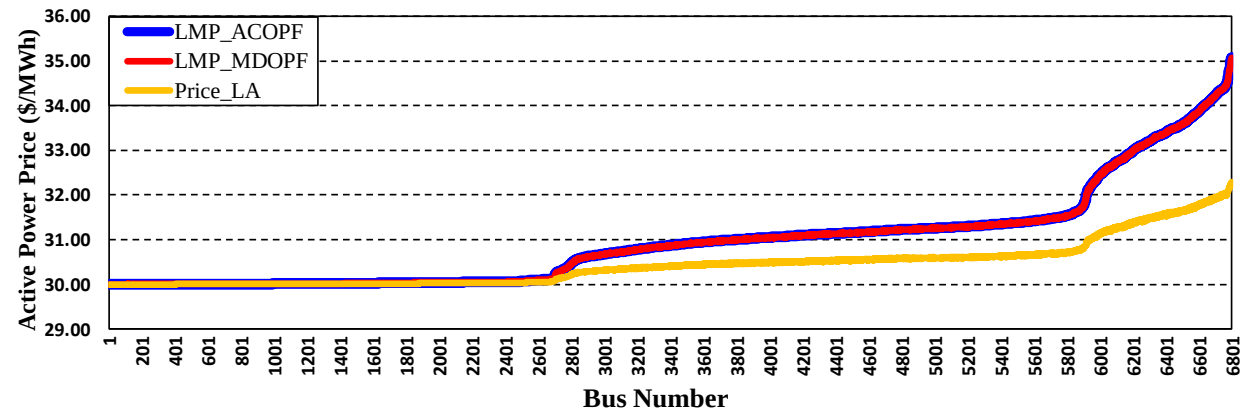

Fig. 7. 6801-bus system.

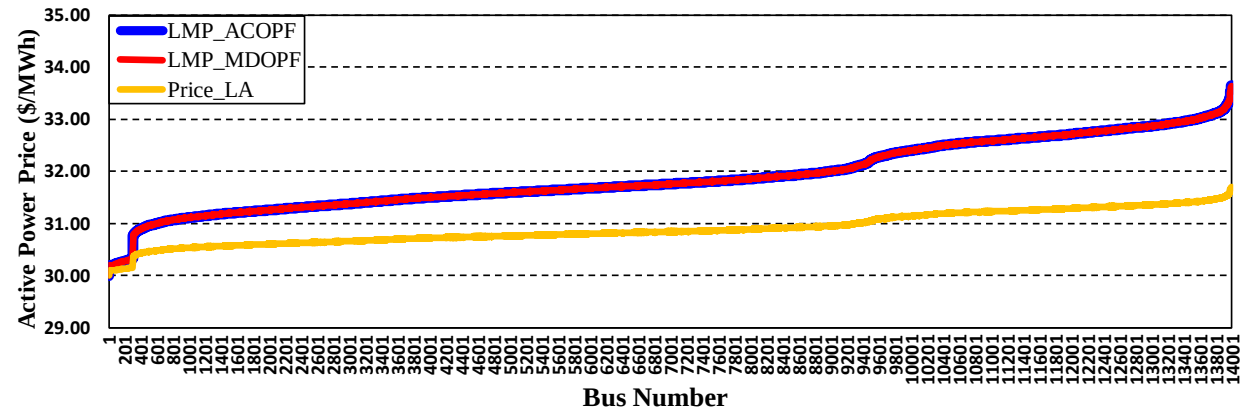

Fig. 8. 14001-bus system.

For all the scenarios, the optimal solutions were successfully solved by MDOPF. Then, the active power prices for different locations were calculated by the proposed MLM and LAM. For comparison, the benchmarking ACOPF results were calculated with MATPOWER. The results are sorted and illustrated in Fig. 6-8. The DLMP errors w.r.t. those of ACOPF are shown in Table III. The OCL by ACOPF and the proposed LAM are shown in Table IV. The solution time of MDOPF is shown in Table V.

As shown in the three figures, the results of MDOPF and the proposed MLM closely match the results of ACOPF. The red price curve always covers the blue curve, even at the end of the distribution feeders, where the prices are relatively higher due to higher losses.

From the errors shown in Table III, we can find that the average errors of active power prices are within a very small range, less than 0.03%, and the largest errors are less than 0.1%. The errors of reactive power prices are slightly larger than those of active power prices because of the small base value (3 $/MVar).

Table IV shows that the OCL caused by ACOPF is significant. Therefore, it is meaningful to develop a price mechanism. By applying DLP and the proposed LAM, the OCL is reduced to approximately 0. Although the LAM cannot ensure that the OCL is 0, it can be finally achieved in practice, where the allocation proportion of losses can be determined by the LAM under the condition that the OCL is known.

TABLE III
SUMMARY OF AVERAGE ERRORS OF DLMP RESULTS

| | 3201-bus system | | 6801-bus system | | 14001-bus system | |
|---|---|---|---|---|---|---|
| | $DLMP_P$ | $DLMP_Q$ | $DLMP_P$ | $DLMP_Q$ | $DLMP_P$ | $DLMP_Q$ |
| Ave. Err. | 0.024% | 0.177% | 0.013% | 0.087% | 0.003% | 0.028% |
| Max. Err. | 0.096% | 0.838% | 0.086% | 0.577% | 0.013% | 0.155% |

TABLE IV
OVER-COLLECTION OF LOSSES WITH THE MLM AND LAM

| | 3201-bus system | 6801-bus system | 14001-bus system |
|---|---|---|---|
| MLM | 405.55 $ | 503.30 $ | 1582.87 $ |
| LAM | -0.82 $ | -0.65 $ | -0.32 $ |

TABLE V
COMPUTATIONAL EFFICIENCY OF MDOPF

| | 3201-bus system | 6801-bus system | 14001-bus system |
|---|---|---|---|
| Solution time | 1.8756 s | 2.6214 s | 4.3360 s |

In addition, Table V shows that the proposed MDOPF can achieve fast computation. This verifies that the proposed model works consistently on different large systems.

In summary, we conclude that the proposed MDOPF is much more accurate than existing benchmarks, DLMP enjoys satisfactory accuracy for large systems, and DLP can eliminate OCL effectively.

## VI. CONCLUSION

In this paper, a convex power flow model named MDOPF is proposed. MDOPF can achieve very similar optimal dispatches of active and reactive power to those of ACOPF. To provide price information for energy consumption, two price mechanisms, namely, DLMP and DLP, are discussed. For the widely considered DLMP, an explicit MLM is developed to improve the accuracy and reduce the computational complexity. Theoretical analysis and numerical tests both show that the MLM will cause OCL. To eliminate OCL, DLP and a novel LAM are proposed. Numerical tests show that this method is effective. As the proposed MLM and LAM have explicit forms, we also look forward to applying them to other ADN issues.

## APPENDIX

*Proof:*

Since $V_0$ is the voltage magnitude of the slack bus, which is fixed in the optimization process, $C_1(\hat{P}_0^G, \hat{Q}_0^G)$ is linear.

According to (47), $V_R^D$ is fixed, so $C_2(\hat{P}_R^G, \hat{Q}_R^G)$ is linear. Therefore, we need to prove that $C_3(\hat{P}_R^G, \hat{Q}_R^G)$ is convex.

$C_3(\hat{P}_R^G, \hat{Q}_R^G)$ can be written in matrix form:

$$\begin{bmatrix} (\hat{P}_R^G)^T & (\hat{Q}_R^G)^T \end{bmatrix} \begin{bmatrix} T^T R_N T & T^T R_N T \\ T^T X_N T & T^T X_N T \end{bmatrix} \begin{bmatrix} C_R^P & 0 \\ 0 & C_R^Q \end{bmatrix} \begin{bmatrix} \hat{P}_R^G \\ \hat{Q}_R^G \end{bmatrix} \quad (94)$$

Let $2\nabla^2 f$ denote the Hessian matrix. $\nabla^2 f$ can be expressed as the Hadamard product of two matrices, i.e., $\nabla^2 f = \nabla^2 f_1 \circ \nabla^2 f_2$, where $\nabla^2 f_1$ and $\nabla^2 f_2$ are:

$$\nabla^2 f_1 = \begin{bmatrix} T^T R_N T & T^T R_N T \\ T^T X_N T & T^T X_N T \end{bmatrix}. \quad (95)$$

$$\nabla^2 f_2 = \begin{bmatrix} C_1^P & \cdots & C_N^P & C_1^Q & \cdots & C_N^Q \\ \vdots & \ddots & \vdots & \vdots & \ddots & \vdots \\ \underbrace{C_1^P \;\; \cdots \;\; C_N^P}_{2N \times N} & & & \underbrace{C_1^Q \;\; \cdots \;\; C_N^Q}_{2N \times N} & & \end{bmatrix}. \quad (96)$$

First, we prove that $\nabla^2 f_1$ is positive semidefinite (PSD):

$$\nabla^2 f_1 = \begin{bmatrix} T^T R_N T & T^T R_N T \\ T^T X_N T & T^T X_N T \end{bmatrix} = \begin{bmatrix} T & 0 \\ 0 & T \end{bmatrix}^T \begin{bmatrix} R_N & R_N \\ X_N & X_N \end{bmatrix} \begin{bmatrix} T & 0 \\ 0 & T \end{bmatrix} \quad (97)$$

Because $T$ is an upper triangular matrix and its diagonal elements are 1, $T$ is invertible. Therefore, $\nabla^2 f_1$ is congruent with the following matrix:

$$\begin{bmatrix} T^T R_N T & T^T R_N T \\ T^T X_N T & T^T X_N T \end{bmatrix} \cong \begin{bmatrix} R_N & R_N \\ X_N & X_N \end{bmatrix} \quad (98)$$

It is apparent that the right-hand side is PSD, so $\nabla^2 f_1$ is PSD.

Additionally, since the elements of each column of $\nabla^2 f_2$ are equal, the rank of $\nabla^2 f_2$ is 1, and the nonzero eigenvalue of $\nabla^2 f_2$ is the trace of $\nabla^2 f_2$. If the trace $\mathrm{tr}(\nabla^2 f_2)$ is positive, $\nabla^2 f_2$ is PSD. In OPF problems, the summation of generation marginal costs is usually larger than zero.

In linear algebra, the Schur product theorem states that the Hadamard product of two PSD matrices is also PSD. Therefore, $\nabla^2 f$ is PSD, and $C_3(\hat{P}_R^G, \hat{Q}_R^G)$ is convex. ■